# Cascades of turbulent kinetic energy and multicomponent scalars in a momentum-scalar coupling turbulence driven by multiple mechanisms under homogeneous and isotropic hypotheses

Wei Zhao

*State Key Laboratory of Photon-Technology in Western China Energy, International Scientific and Technological Cooperation Base of Photoelectric Technology and Functional Materials and Application, Laboratory of Optoelectronic Technology of Shaanxi Province, Institute of Photonics and Photon-technology, Northwest University, Xi'an 710127, China*

Correspondence: zwbayern@nwu.edu.cn

**Abstract**

Momentum-scalar coupling turbulence, a phenomenon observed in both natural and engineering contexts, involves the intricate interaction between multicomponent scalars and multiscale forces (i.e. multiple coupling mechanisms), resulting in a wide array of manifestations. Despite its importance, limited research has been conducted to comprehend the influence of these multicomponent and multiple coupling mechanisms on turbulence cascades. Hence, this study aims to provide a preliminary and theoretical exploration into how these multiple coupling mechanisms govern the cascades of turbulent kinetic energy and multicomponent scalars. To simplify the mathematical analysis, homogeneous and isotropic hypotheses of flow field have been applied. The key findings of this study can be summarized as follows: (1) Validation of Quad-cascade processes. (2) Examination of various cases involving single scalar components but multiple coupling mechanisms. Of particular interest is the coexistence of buoyancy-driven turbulence and electrokinetic turbulence, which introduces a new variable fluxes (VF) subrange resulting from their nonlinear interaction. Another extension considers an exponential modulation function, equivalent to the coexistence of multiple coupling mechanisms acting on a single scalar. The study identifies two new VF subranges. (3) Binary scalar components and coupling mechanisms are investigated, indicating coupling mechanisms with significantly different strengths can also induce complex interactions and new VF subranges. (4) Complexity when three or more different scalar components and coupling mechanisms coexist simultaneously: With the exception of certain special cases, closure of the problem becomes unattainable. This highlights the challenges inherent in addressing the simultaneous presence of multiple scalar components and coupling mechanisms. This research endeavor illuminates the theoretical understanding of the diverse scaling properties observed in momentum-scalar coupling turbulence across different scenarios.

**1. Introduction**

Turbulence is pervasive in nature and engineering, arising from various physical mechanisms such as hydrodynamics, thermal convection (Jiang et al. 2020), electrohydrodynamics (Zhang & Zhou 2023), magnetohydrodynamics (Knaepen & Moreau 2008, Eyink et al. 2013), and even in quantum (Madeira et al. 2020, Mäkinen et al. 2023) and biological (Alert et al. 2022) systems. Turbulent transport involves diverse scalars such as temperature (Niemela et al. 2000, Ahlers et al. 2009), salinity (Davidson 2013, Fan et al. 2018), electric conductivity (Wang et al. 2014, Wang et al. 2016, Wang et al. 2016), and permittivity (Varshney et al. 2016, Zhao & Wang 2019), among others like magnetic susceptibility (Kitazawa et al. 2001) and chemical components (Kuramoto 1980, Pope 1985). These scalars display a multiscale distribution in wavenumber space and can generate external multiscale forces through different physical mechanisms, driving the flow towards a state of turbulence through a feedback mechanism.

In the real world, it's rare for just a single scalar or physical mechanism to be isolated. Instead, the coexistence of multicomponent scalars and multiphysical interactions is the normal state. In the atmosphere, for example, buoyancy-driven flow, electrohydrodynamic flow due to spatial electric charge distribution (Kikuchi 2001, Zhang & Zhou 2023), and magnetohydrodynamic flow influenced by the Earth's magnetic field (El-Alaoui et al. 2021) all occur concurrently. Buoyancy-driven turbulence—which includes thermal convection, stratified turbulence, and boundary layers—is a key mechanism in atmospheric evolution on scales up to hundreds of kilometers. The atmosphere contains various electric media such as charged species, gases with differing moisture levels, droplets, and dust clouds, leading to spatial and temporal variations in electric conductivity and permittivity. Given that the atmosphere is not electrically neutral and can experience electric fields as strong as $10^6$ V/m during thunderstorms, electrohydrodynamic (EHD) or electrokinetic (EK) turbulence—both driven by electric body force (EBF)—are also inevitable in atmosphere flow up to kilometer scales. Additionally, whenever charged species are unevenly distributed, deviations from electrical neutrality arise locally. This scenario leads to the flow carrying nonuniform charges, resulting in electric currents. Under the Earth's magnetic field, these currents produce Lorentz forces (LF), necessitating the presence of magnetohydrodynamic (MHD) effects. Thus, buoyancy-driven turbulence, EBF-driven turbulence, and LF-driven (MHD) turbulence coexist and can collectively dictate wind evolution in both large and small scales.

Similar examples of turbulence driven by multiscale forces exist elsewhere. In the ocean, temperature and salinity jointly affect flows on various scales. At the core of Earth (He et al. 2022), the combination of high and uneven temperatures with ferromagnetism can mutually impact the dynamics of the outer liquid core. Moreover, in fields like chemical engineering and combustion, multicomponent scalar systems are also prevalent (Perry & Mueller 2018). These scenarios highlight the intricate interplay and importance of understanding multiscalar and multiphysical interactions in turbulent systems.

Unfortunately, most theoretical and numerical researches study the "real world turbulence" (Lesieur 2008) by focusing on a single scalar and one dominant physical mechanism, such as buoyancy or electrostatic influence. A paradigm of momentum-scalar coupled turbulent flow—buoyancy-driven turbulence, has attracted significant attention over the past several decades. Obukhov's work (Obukhov 1949) laid the theoretical groundwork for understanding the inertial subrange of passive scalar transport in turbulence characterized by high Reynolds numbers. Subsequently, Bolgiano (1959) and Obukhov (1959) independently introduced the Bolgiano-Obukhov law (BO59) (Lohse & Xia 2010). According to their predictions, in the subrange primarily influenced by buoyancy effects arising from density or temperature variations, the spectra of turbulent kinetic energy and scalar variance demonstrate specific characteristics as

$$E_u(k) \sim k^{-\frac{11}{5}} \quad \text{and} \quad E_s(k) \sim k^{-\frac{7}{5}} \tag{1}$$

The buoyancy-dominant subrange described by BO59 law represents a region characterized by partially variable flux, within this context, the flux of scalar variance remains constant while that of turbulent kinetic energy does not.

Since the influence of buoyancy decays rapidly along wavenumber, the buoyancy-dominant subrange occurs at the lower wavenumber side of the inertial subrange. In 2000, Niemela et al. (2000) first observed slopes of -11/5 and -7/5 in the spectra of turbulent kinetic energy and temperature variance, respectively. Subsequently, several researchers (Kumar et al. 2014, Kumar & Verma 2015, Wang et al. 2022) claimed that the BO59 law can be predicted through numerical simulations across various scenarios. However, both numerical and experimental investigations still lack convincing results regarding the BO59 law. For a more comprehensive review of buoyancy-driven turbulence, please refer to Verma et al (2017). In an effort to address this issue, Alam et al. (2019) revisited the BO59 law and attempted to explain why its observation has not been successful. Notably, their work provided numerical observations of a novel scaling subrange featuring nonconstant flux of scalar variance and constant flux of kinetic energy, which, in my opinion, constitutes their most significant contribution. Regrettably, this observation did not receive much attention at the time.

Electrokinetic turbulence pertains to a turbulence phenomenon primarily observed in electrolyte solutions with nonuniform electric properties (such as electric conductivity or permittivity), driven by EBF through electrostatic effects. Regarding EK turbulence, Zhao and Wang (2017, 2019) have developed a series of theoretical frameworks aimed at elucidating the transport of turbulent kinetic energy and scalar variance. Based on a hypothesis involving a constant flux of scalar variance (electric conductivity or permittivity) while nonconstant flux of kinetic energy, they predicted that the spectra of turbulent kinetic energy and scalar variance would exhibit

$$E_u(k) \sim k^{-\frac{7}{5}} \quad \text{and} \quad E_s(k) \sim k^{-\frac{9}{5}} \tag{2}$$

Furthermore, it was predicted that in EK turbulence, where the EBF becomes dominant at small scales, the subrange where the cascades of turbulent kinetic energy and scalar variance are influenced by the EBF should occur at the higher wavenumber range of the inertial subrange. As a result, EK turbulence is characterized by a bidirectional cascade of turbulent kinetic energy and a direct cascade of scalar variance (Zhao et al. 2021). This represents another significant difference between buoyancy-driven turbulence and EK turbulence.

In magnetohydrodynamic turbulence, numerous theories have been proposed over the past 60 years. Iroshinikov (1964) and Kraichnan (1965) separately introduced the Iroshinikov-Kraichnan spectrum for MHD turbulence by considering the influence of the Alfvén effect on the cascade of turbulent kinetic energy (Biskamp 2003). According to their theory, the energy spectrum $E_u(k)$ in the inertial subrange of MHD turbulence follows a power law of $k^{-3/2}$, deviating from the classical Kolmogorov -5/3 law, even though the latter is more commonly observed in experimental data (Leamon et al. 1998, Chepurnov & Lazarian 2010) and numerical simulations of MHD turbulence (Biskamp 2003, Verma 2004). In contrast, the Iroshinikov-Kraichnan theory was numerically supported by Eyink et al (2013) when considering a Richardson dispersion, besides the Alfvén effect. Further theoretical models, such as the anisotropic MHD energy spectrum $E_u(k) \sim k^{-5/2}$ have also been proposed for the inertial subrange of anisotropic MHD turbulence (Biskamp 2003). Moreover, experimental investigations (El-Alaoui et al. 2021) have revealed a wider array of scaling behaviors. Despite these advances, the research focus has largely been on the form of the inertial subrange, not so much on the more intricate subrange where the Lorentz force takes precedence in the cascades of turbulent kinetic energy and related scalar fields like magnetic conductivity. Verma (2022) provides an extensive overview of recent developments in this field.

As can be seen above, the current discourse on turbulence often follows a reductionist approach, suggesting that a single primary mechanism largely governs these phenomena, which can overlook the inherent complexity of such systems. In the physical and life sciences (Strogatz et al. 2022), the concept of emergence acknowledges that multifaceted, highly nonlinear systems exhibit behaviors that cannot be attributed solely to one dominant mechanism, ignoring the distinction between primary and secondary effects. This concept, highlighted in Philip Anderson's seminal paper "More is Different" (Anderson 1972), emphasizes the novelty arising from complex interactions across different scales.

Turbulence is a prime example of a highly nonlinear system where isolating individual components and coupling mechanisms is often impractical, especially in fields like engineering, meteorology, oceanography, and astrophysics. In EK or electrohydrodynamic (EHD) turbulence, the interplay of external electric fields with fluids produces both electrostatic forces and electrothermal effects, the latter leading to buoyancy through temperature variations. Traditionally, electrostatically induced EK turbulence is considered the primary factor. However, the influence of electrothermal effects on overall turbulence is not well understood. The presence of such multidimensional coupling mechanisms can lead to unexpected and novel emergent behaviors.

Despite the significance of this multifaceted interplay in turbulent systems, research on the cascades of turbulent kinetic energy and scalar fields influenced by multiple mechanisms is lacking. This study aims to explore the interplay among the physical mechanisms governing the cascade processes of turbulent kinetic energy and multicomponent scalars. It reveals that interactions between coupling mechanisms of distinct strengths can generate new scaling behaviors in the spectra of turbulent kinetic energy and the variances of two scalar quantities.

**2. Theory**

Most theoretical models to date have focused on the inertial subrange (e.g., Kolmogorov's 1941 law, Obukhov-Corrsin law), the constant scalar flux subrange (e.g., BO59 law, Zhao and Wang model (2017, 2019, 2021)), and the variable flux (VF) subrange (e.g., Verma (2022), Shur-Lumley model (Basu & Holtslag 2022)). In 2022, inspired by Alam et al.'s investigation of stably stratified turbulence (2019), Zhao (2022) developed a model for momentum-scalar coupling turbulence based on the fluxes of turbulent kinetic energy ($\Pi_u$) and scalar variance ($\Pi_s$).

This model employed a multiscale force of $\nabla^a s'$-type (also see (Zhao & Wang 2021)) to describe the coupling relationship between momentum and scalar fluctuations ($s'$). It theoretically predicted a quad-cascade process involving turbulent kinetic energy and scalar variance, covering the inertial subrange, constant energy flux (constant-$\Pi_u$) subrange, constant scalar flux (constant-$\Pi_s$) subrange, and VF subrange. Zhao's model (Zhao 2022) provides a comprehensive framework for studying complex momentum-scalar coupling turbulence with multiple scalar components and coupling mechanisms, encompassing various phenomena such as buoyancy-driven flow, electrokinetic flow, and magnetohydrodynamic flow with differing $a$ values.

*2.1 Conservative law by the fluxes of turbulent kinetic energy and multicomponent scalars*

In this investigation, based on Zhao's model (Zhao 2022), a flux model for momentum-scalar coupling turbulence with multicomponent scalars and multiple coupling mechanisms is studied theoretically and numerically. A three-dimensional model has been considered. The governing equations are expressed as follows:

$$\frac{D\boldsymbol{u}}{Dt} = -\frac{1}{\rho}\nabla p + \nu\Delta\boldsymbol{u} + \sum_{i=1}^{imax} \boldsymbol{M}_i \mathfrak{D}^{\frac{\beta_i}{4}} s'_i \qquad (3a)$$

$$\begin{cases} \dfrac{Ds_1'}{Dt} = -\boldsymbol{N_1} \cdot \boldsymbol{u} + D_{f,1}\Delta s_1' \\ \quad\vdots \\ \dfrac{Ds_i'}{Dt} = -\boldsymbol{N_i} \cdot \boldsymbol{u} + D_{f,i}\Delta s_i' \\ \quad\vdots \\ \dfrac{Ds_{imax}'}{Dt} = -\boldsymbol{N_{imax}} \cdot \boldsymbol{u} + D_{f,imax}\Delta s_{imax}' \end{cases} \quad (3b)$$

$$\nabla \cdot \boldsymbol{u} = 0 \quad (3c)$$

where $\rho$ is the referenced fluid density, $\boldsymbol{u}$ denotes the velocity vector, $p$ is pressure, $s_i'$ is the fluctuation of the $i^{th}$ scalar component. $D/Dt = d/dt + \boldsymbol{u}\cdot\nabla$ is material derivative. In this manuscript, $\mathfrak{D} = \Delta^2$, with $\Delta = \nabla\cdot\nabla$ being Laplacian operator. $\mathfrak{D}^{\beta_i/4} s_i'$ denote the fractional derivation on $s_i'$ with orders $\beta_i/4$ respectively. $\boldsymbol{M_i}\mathfrak{D}^{\beta_i/4}s_i'$ is the multiscale force related to the $i$th scalar field $s_i'$ (Zhao & Wang 2021, Zhao 2022), with $\boldsymbol{M_i}$ being the dimensional vector associated with the $i$th physical field. $\boldsymbol{N_i} = \nabla \bar{s_i}$ is the gradient of the mean field ($\bar{\ }$) with respect to the $i$th scalar $s_i$. It characterizes the strength of the mean scalar gradient that can be provided from the background flow. $\nu$ and $D_{f,i}$ are the kinematic viscosity and diffusivity of the $i$th scalar respectively. The assumption ($\nabla^2 \bar{s_i} \ll \nabla^2 s_i'$) is made for the model.

For example, in buoyancy-driven turbulence, it is approximately $\beta_i = 0$. $s_i'$ can be either temperature fluctuations with $\boldsymbol{M_i} = \alpha g\hat{\boldsymbol{z}}$ and $\boldsymbol{N_i} = \nabla\bar{T}$, or density fluctuations with $\boldsymbol{M_i} = \boldsymbol{N_i} = -f_{VB}\hat{\boldsymbol{z}}$, respectively. Here, $\alpha$ is thermal expansion coefficient, $g$ is gravity acceleration, $\hat{\boldsymbol{z}}$ is the inverse direction of gravity, $\bar{T}$ is the mean field of temperature ($T$), $f_{VB} = \sqrt{\dfrac{g}{\bar{\rho}}\left|\dfrac{d\rho_0}{dz}\right|}$ is Väisälä–Brunt frequency. $\rho_0$ and $\bar{\rho}$ are background density and mean density respectively (Alam et al. 2019, Verma 2022). In an electrokinetic turbulence generated by electrostatic force, when the electric field ($E$) is in $y$-direction ($\hat{\boldsymbol{y}}$) that perpendicular to initial interface of electric conductivity ($\sigma$), as investigated by Wang et al (2014, 2016), it is approximately $\beta_i = 1$, with $\boldsymbol{M_i} = -\varepsilon E^2 \hat{\boldsymbol{y}}/\rho\bar{\sigma}$ and $\boldsymbol{N_i} = \nabla\bar{\sigma}$. Here, $\varepsilon$ is electric permittivity and $\bar{\sigma}$ is the mean field of electric conductivity. (Zhao 2022)

In this model, two hypotheses have been made to simplify the analysis.

(1) All scalar components are assumed to be independent and non-reactive. This is analogous to the well-known Boussinesq approximation (Zeytounian 2003, Verma et al. 2017). For instance, in buoyancy-driven turbulence, buoyancy can result from stratifications of temperature, density (different materials), and salinity, which can be treated as independent of each other. Similarly, in electrokinetic turbulence in an electrolyte solution with different free ions (e.g., $Na^+$, $Ca^{2+}$, $Cl^-$), ions are considered independent and non-reactive.

(2) Turbulent flow is approximated as homogeneous and isotropic. This approximation is common in many classical turbulent models, such as the Kolmogorov 1941 law, Obukhov-Corrsin law, and BO59 law for stably stratified turbulence, as well as the Zhao-Wang model (2017, 2019) for electrokinetic turbulence. This hypothesis is partially supported by studies such as Nath et al. (2016) in turbulent thermal convection and Kumar et al. (2014) in stably stratified turbulence.

In Fourier space, let $E_u(\boldsymbol{k}) = \tfrac{1}{2}|\boldsymbol{u}(\boldsymbol{k})|^2$ and $E_{s,i}(\boldsymbol{k}) = \tfrac{1}{2}|s_i'(\boldsymbol{k})|^2$ are the modal turbulent kinetic energy and $i$th scalar variance, Eqs. (3) can be rewritten as (Davidson 2004, Lesieur 2008, Verma 2022)

$$\dfrac{d}{dt}E_u(\boldsymbol{k}) = T_u(\boldsymbol{k}) - D_u(\boldsymbol{k}) + \sum_{i=1}^{imax} F_{s,i}(\boldsymbol{k}) \quad (4a)$$

$$\frac{d}{dt}E_{s,i}(\boldsymbol{k}) = T_{s,i}(\boldsymbol{k}) - D_{s,i}(\boldsymbol{k}) - F_{A,i}(\boldsymbol{k}) \tag{4b}$$

$$k_l u_l(k) = 0 \tag{4c}$$

with

$$T_u(\boldsymbol{k}) = \text{Im}\left[\int_{\square}^{\square} k_l \widehat{u_l}(n) \widehat{u_q}(m) \widehat{u_q^*}(k) dk^3\right] \tag{5a}$$

$$T_{s,i}(\boldsymbol{k}) = \text{Im}\left[\int_{\square}^{\square} k_l \widehat{u_l}(n) \widehat{s_i'}(m) \widehat{s_i'^*}(k) dn^3\right] \tag{5b}$$

$$F_{s,i}(\boldsymbol{k}) = \text{Re}\left[k^{\beta_i} M_{i_q} \widehat{s_i'}(k) \widehat{u_q^*}(k)\right] \tag{5c}$$

$$F_{A,i}(\boldsymbol{k}) = \text{Re}\left[N_{i_q} \widehat{s_i'}(k) \widehat{u_q^*}(k)\right] \tag{5d}$$

$$D_u(\boldsymbol{k}) = 2\nu k^2 E_u(\boldsymbol{k}) \tag{5e}$$

$$D_{s,i}(\boldsymbol{k}) = 2 D_{f,i} k^2 E_{s,i}(\boldsymbol{k}) \tag{5f}$$

where Re and Im represent the real and imaginary parts of the quantity. $\boldsymbol{k} = k_l \widehat{x_l}$ and $k = |\boldsymbol{k}| = (k_l k_l)^{1/2}$, with $k_l$ being the wavenumber component in the $l$th direction (denoted by $\widehat{x_l}$). The wavenumbers have a relation of $\boldsymbol{k} = \boldsymbol{m} + \boldsymbol{n}$. $\widehat{u_q}$ is the Fourier transform of the $q$th directional component of $\boldsymbol{u}$. $\widehat{s_i'}$ is the Fourier transform of $s_i'$. The asterisk represents a complex conjugate. $T_u(\boldsymbol{k})$ and $D_u(\boldsymbol{k})$ are the nonlinear turbulent kinetic energy transfer rate and dissipation rate, respectively. $T_{s,i}(\boldsymbol{k})$ and $D_{s,i}(\boldsymbol{k})$ are the nonlinear transfer rate of the $i$th scalar variance and scalar dissipation rate, respectively. $F_{s,i}(\boldsymbol{k})$ denotes the energy feeding rate by the multiscale force due to the $i$th scalar field and $F_{A,i}(\boldsymbol{k})$ is the $i$th scalar feeding rate by bulk components.

Assume the velocity and scalar fields are homogeneous and isotropic, in a spherical shell around $k$ with a thickness of $dk$ in the wavenumber space, Eqs. (4a-b) become (Verma 2018)

$$\sum_{k<|\boldsymbol{k}'|\leq k+dk} \frac{d}{dt} E_u(\boldsymbol{k}') = \sum_{k<|\boldsymbol{k}'|\leq k+dk}\left[T_u(\boldsymbol{k}') + \sum_{i=1}^{imax} F_{s,i}(\boldsymbol{k}') - D_u(\boldsymbol{k}')\right] \tag{6a}$$

$$\sum_{k<|\boldsymbol{k}'|\leq k+dk} \frac{d}{dt} E_{s,i}(\boldsymbol{k}') = \sum_{k<|\boldsymbol{k}'|\leq k+dk}\left[T_{s,i}(\boldsymbol{k}') - F_{A,i}(\boldsymbol{k}') - D_{s,i}(\boldsymbol{k}')\right] \tag{6b}$$

Since the flux of turbulent kinetic energy $\Pi_u(k) = -\sum_{|\boldsymbol{k}'|\leq k} T_u(\boldsymbol{k}')$ and the flux of scalar variance $\Pi_{s,i}(k) = -\sum_{|\boldsymbol{k}'|\leq k} T_{s,i}(\boldsymbol{k}')$ (Alam et al. 2019, Verma 2022, Zhao 2022), we can define

$$d\Pi_u(k) = -\sum_{k\leq|\boldsymbol{k}'|\leq k+dk} T_u(\boldsymbol{k}') \tag{7a}$$

$$d\Pi_{s,i}(k) = -\sum_{k\leq|\boldsymbol{k}'|\leq k+dk} T_{s,i}(\boldsymbol{k}') \tag{7b}$$

$$E_u(k) dk = \sum_{k<|\boldsymbol{k}'|\leq k+dk} E_u(\boldsymbol{k}') \tag{7c}$$

$$E_{s,i}(k)dk = \sum_{k<|\mathbf{k'}|\le k+dk} E_{s,i}(\mathbf{k'}) \tag{7d}$$

$$F_{s,i}(k)dk = \sum_{k<|\mathbf{k'}|\le k+dk} F_{s,i}(\mathbf{k'}) \tag{7e}$$

$$F_{A,i}(k)dk = \sum_{k<|\mathbf{k'}|\le k+dk} F_{A,i}(\mathbf{k'}) \tag{7f}$$

$$D_u(k)dk = \sum_{k<|\mathbf{k'}|\le k+dk} D_u(\mathbf{k'}) \tag{7g}$$

$$D_{s,i}(k)dk = \sum_{k<|\mathbf{k'}|\le k+dk} D_{s,i}(\mathbf{k'}) \tag{7h}$$

where $E_u(k)$ and $E_{s,i}(k)$ are the averaged power spectra of turbulent kinetic energy and the $i$th scalar variance over the spherical shell, $F_{s,i}(k)$ is the energy feeding rate by multiscale force due to the $i$th scalar field at wavenumber $k$, $F_{A,i}(k)$ is the $i$th scalar feeding rate at wavenumber $k$, $D_u$ and $D_{s,i}$ are dissipation terms at wavenumber $k$ respectively. After substituting Eqs. (7a-h) into Eqs. (6a-b), considering the flow is statistically equilibrium, and let $dk \to 0$, it is obtained

$$\frac{d}{dk}\Pi_u(k) = \sum_{i=1}^{imax} F_{s,i}(k) - D_u(k) \tag{8a}$$

$$\frac{d}{dk}\Pi_{s,i}(k) = -F_{A,i}(k) - D_{s,i}(k) \tag{8b}$$

where

$$F_{s,i}(k) = \text{Re}\left[k^{\beta_i} M_{i_q} \widehat{s_i}(k)\widehat{u_q^*}(k)\right] \tag{9a}$$

$$F_{A,i}(k) = \text{Re}\left[N_{i_q} \widehat{s_i}(k)\widehat{u_q^*}(k)\right] \tag{9b}$$

$$D_u(k) = 2\nu k^2 E_u(k) \tag{9c}$$

$$D_{s,i}(k) = 2D_{f,i} k^2 E_{s,i}(k) \tag{9d}$$

Thus, if $\mathbf{M}_i$ is parallel to $\mathbf{N}_i$,

$$F_{s,i}(k) - \frac{M_i}{N_i} F_{A,i}(k) k^{\beta_i} = 0 \tag{10}$$

where $M_i = |\mathbf{M}_i|$ and $N_i = |\mathbf{N}_i|$ respectively.

*2.2 Inertial subrange*

In the inertial subrange, where the influence of multiscale forcing and dissipation is negligible, Eqs. (8a-b) become

$$\frac{d}{dk}\Pi_u(k) = 0 \tag{11a}$$

$$\frac{d}{dk}\Pi_{s,i}(k) = 0 \tag{11b}$$

Or

$$\Pi_u(k) = \text{const along } k$$

$$\Pi_{s,i}(k) = \text{const along } k$$

In the flow region far from boundary layers, known as the bulk region, the statistical features can significantly differ from those in the boundary layers, as seen in buoyancy-driven turbulence (Verma et al. 2017). In the bulk region, the following relationship can be established (Alam et al. 2019, Zhao 2022)

$$E_u(k) = u_k^2/k \sim k^{\xi_u} \tag{12a}$$

$$E_{s,i}(k) = s_{k,i}^2/k \sim k^{\xi_{s,i}} \tag{12b}$$

$$\Pi_u(k) = k u_k^3 \sim k^{\lambda_u} \tag{12c}$$

$$\Pi_{s,i}(k) = k s_{k,i}^2 u_k \sim k^{\lambda_{s,i}} \tag{12d}$$

where $\xi_u, \xi_{s,i}, \lambda_u, \lambda_{s,i}$ denote the scaling exponents of $E_u, E_{s,i}, \Pi_u$ and $\Pi_{s,i}$ respectively. $u_k$ and $s_{k,i}$ represent the velocity and $i$th scalar components in $k$-space. It can be deduced that

$$E_u^{3/2} k^{5/2} = \text{const along } k \tag{13a}$$

$$E_{s,i} E_u^{1/2} k^{5/2} = \text{const along } k \tag{13b}$$

Thereafter, the celebrated K41 law and Obukhov-Corrsin law for inertial subrange have been derived, with $\xi_u = \xi_{s,i} = -5/3$. Considering Eqs. (11a) is a direct consequence of Navier-Stokes equation, if Navier-Stokes equation has smooth and power-law solutions in spectral space, the $k^{-5/3}$ spectrum related to K41 law must be one of them. From these equations, it can be inferred that in the inertial subrange, there must be constant $\Pi_u$ and $\Pi_{s,i}$ for all the scalars, which are independent of each other. Each scalar will exhibit a -5/3 slope in the spectra of scalar variance, although with different spectral intensities.

*2.3 Conservative law in the MFD subrange*

From Eq. (9a), it is evident that the multiscale force depends not only on the strong coupling between the scalar and velocity spectra, but is also modulated by $k^{\beta_i}$, where $\beta_i$ is primarily determined by the physical mechanism. The influence of the work input by the multiscale force extends across a wide subrange in the spectral space, rather than being limited to a single wavenumber (or frequency) or a narrow spectral band. This is why the forcing term in Eq. (3a) is referred to as a multiscale force. In the subrange where the transports of turbulent kinetic energy and scalar variance are dominated by the multiscale force, known as the multiscale-force dominated (MFD) subrange, the dissipation terms of turbulent kinetic energy and scalar variance are neglected. As a result, Eqs. (8a-b) simplify to

$$\frac{d}{dk}\Pi_u(k) = \sum_{i=1}^{imax} F_{s,i}(k) \tag{14a}$$

$$\frac{d}{dk}\Pi_{s,i}(k) = -F_{A,i}(k) \tag{14b}$$

Combining Eqs. (10) and (14a-b), a conservation law is thus obtained

$$\frac{d}{dk}\Pi_u(k) + \sum_{i=1}^{imax} \frac{M_i}{N_i} k^{\beta_i} \frac{d}{dk}\Pi_{s,i}(k) = 0 \tag{15}$$

It can be seen, Eqs. (11a-b) are the special solutions of Eq. (15) as well. Therefore, the scenarios of Kolmogorov (Kolmogorov 1941, Frisch 1995), Obukhov and Corrsin (Obukhov 1949, Corrsin 1951) theories have been unified into this model. After substituting Eq. (12c-d) into Eq. (15), it is obtained

$$\frac{d}{dk} k u_k^3 + \sum_{i=1}^{imax} \frac{M_i}{N_i} k^{\beta_i} \frac{d}{dk} k s_{k,i}^2 u_k = 0 \tag{16}$$

If $u_k$ is nontrivial, Eq. (16) can be rewritten by multiplying $u_k$ as

$$u_k \frac{d}{dk} k u_k^3 + k u_k^2 \sum_{i=1}^{imax} \frac{M_i}{N_i} k^{\beta_i} \frac{d}{dk} s_{k,i}^2 + \left(\sum_{i=1}^{imax} \frac{M_i}{N_i} k^{\beta_i} s_{k,i}^2\right) u_k \frac{d}{dk} k u_k = 0 \tag{17}$$

Given that in the MFD subrange, the flow is propelled by the $\sum_{i=1}^{imax} M_i \mathfrak{D}^{\beta_i/4} s_i'$ type force, as expressed in Eq. (3a). Thus, another relationship that connects $u_k$ and $s_{k,i}$ is established dimensionally as

$$k u_k^2 = \sum_{i=1}^{imax} k^{\beta_i} M_i s_{k,i} \tag{18}$$

Subsequently, by substituting Eq. (18) into (17) with simple mathematical processing, it is obtained

$$\sum_{j=1}^{imax} \sum_{i=1}^{imax} M_i M_j \left[ \frac{3}{2} k^{\beta_i+\beta_j-1} s_{k,i} \left(\frac{d}{dk} s_{k,j}\right) + \frac{1}{2}(3\beta_j - 1) k^{\beta_i+\beta_j-2} s_{k,i} s_{k,j} + \frac{2}{N_j} k^{\beta_i+\beta_j} s_{k,i} s_{k,j} \left(\frac{d}{dk} s_{k,j}\right) + \right.$$
$$\left. \frac{\beta_j+1}{2N_i} k^{\beta_i+\beta_j-1} s_{k,j} s_{k,i}^2 + \frac{1}{2N_i} k^{\beta_i+\beta_j} s_{k,i}^2 \left(\frac{d}{dk} s_{k,j}\right) \right] = 0 \tag{19}$$

Since $k^{\beta_i+\beta_j-2} s_{k,i}$ is nontrivial, finally Eq. (19) becomes

$$\sum_{j=1}^{imax} \sum_{i=1}^{imax} F_{ij} = 0 \tag{20}$$

with

$$F_{ij} = M_i M_j \left[ \left(\frac{3}{2} k + \frac{2}{N_j} k^2 s_{k,j} + \frac{1}{2N_i} k^2 s_{k,i}\right) \left(\frac{d}{dk} s_{k,j}\right) + \frac{1}{2}(3\beta_j - 1) s_{k,j} + \frac{\beta_j+1}{2N_i} k s_{k,i} s_{k,j} \right] \tag{21}$$

$F_{ij}$ represents the interaction between the *i*th scalar and its corresponding coupling mechanism with the *j*th one. In a turbulent system with $imax \geq 2$, e.g. scalar A and B or their coupling mechanisms, altering their indices does not alter the underlying physics or observations. In other words, the cascades of turbulent kinetic energy and scalar variances remain unaffected. Consequently, a corollary is proposed as follows.

**Corollary**: $F_{ij}$ is invariant when switching $i$ and $j$, i.e. $F_{ij} = F_{ji}$, accordingly

$$\frac{3}{2} k \frac{d}{dk}(s_{k,j} - s_{k,i}) + k^2 \frac{d}{dk}\left(\frac{s_{k,j}^2}{N_j} - \frac{s_{k,i}^2}{N_i}\right) + \left(\frac{1}{2N_i} k^2 s_{k,i} \frac{d}{dk} s_{k,j} - \frac{1}{2N_j} k^2 s_{k,j} \frac{d}{dk} s_{k,i}\right)$$
$$+ \frac{3}{2}(\beta_j s_{k,j} - \beta_i s_{k,i}) - \frac{1}{2}(s_{k,j} - s_{k,i}) + \frac{k s_{k,i} s_{k,j}}{2 N_i N_j}(\beta_j N_j + N_j - \beta_i N_i - N_i) = 0 \tag{22}$$

Then, the computation on $s_{k,i}$ can be simplified according to this symmetry.

## 2.4 Numerical computations

To demonstrate the variations of $E_u(k)$, $E_{s,i}(k)$, $\Pi_u(k)$ and $\Pi_{s,i}(k)$, a wide range of wavenumbers is required. However, direct computation of Eqs. (18), (20), and (21) in a linear wavenumber space with a wide band, e.g. 50 decades, demands excessive computational resources. Therefore, in this investigation, these computations are performed in a nonlinear wavenumber space by transforming $k = 10^q$, as suggested by Zhao (2022). Considering $ds_{k,i}/dk = (10^q \ln 10)^{-1} ds_{k,i}/dq$, Eqs. (18), (20), and (21) are converted to

$$F_{ij} = M_i M_j \left[ (\ln 10)^{-1} \left( \frac{3}{2} + \frac{2}{N_j} 10^q s_{k,j} + \frac{1}{2N_i} 10^q s_{k,i} \right) \frac{ds_{k,i}}{dq} + \frac{1}{2}(3\beta_j - 1)s_{k,j} + \frac{\beta_j + 1}{2N_i} 10^q s_{k,i} s_{k,j} \right] \quad (23)$$

$$u_k^2 = \sum_{i=1}^{imax} 10^{(\beta_i - 1)q} M_i s_{k,i} \quad (24)$$

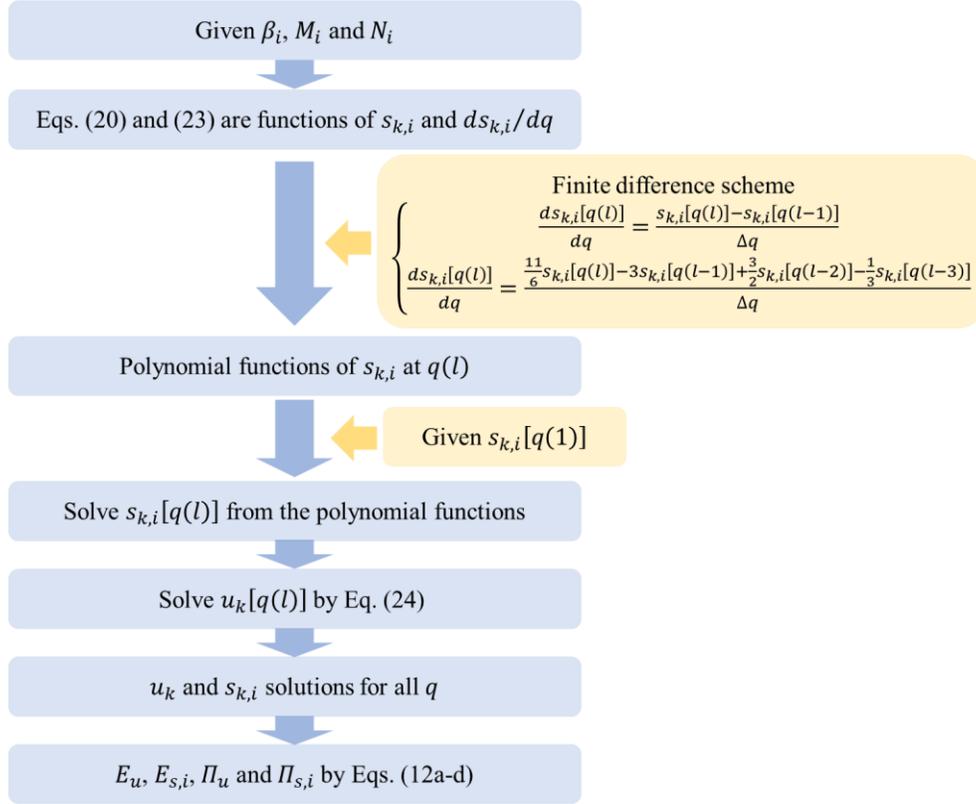

Figure 1. Flow chart of the procedure in solving $s_{k,i}$, $u_k$, and the corresponding $E_u$, $E_{s,i}$, $\Pi_u$ and $\Pi_{s,i}$.

Eqs. (20), (23) and (24) (if solvable, see section 3.4 for details) can be numerically solved using finite difference method, as depicted in Figure 1. During the computation, the process of $ds_{k,i}/dq$ term is crucial. Two finite difference schemes have been applied. At a position $q(l)$ with $l > 1$, a first-order approximation was applied with $ds_{k,i}[q(l)]/dq = \{s_{k,i}[q(l)] - s_{k,i}[q(l-1)]\}/\Delta q$, where $\Delta q = q(l) - q(l-1) = 0.006$ is the interval in the $q$-space. When $l \gg 1$, a third-order approximation is applied, with $ds_{k,i}[q(l)]/dq = \{\frac{11}{6}s_{k,i}[q(l)] - 3s_{k,i}[q(l-1)] + \frac{3}{2}s_{k,i}[q(l-2)] - \frac{1}{3}s_{k,i}[q(l-3)]\}/\Delta q$. Since all the $s_{k,i}[q(l-1)]$, $s_{k,i}[q(l-2)]$, and $s_{k,i}[q(l-3)]$ have been solved in previous steps, substituting $ds_{k,i}[q(l)]/dq$ into Eqs. (20) and (23) leaves $s_{k,i}[q(l)]$ as the only unknowns. Eq. (23) can be solved based on boundary conditions, $s_{k,i}[q(1)]$. Consequently, $u_k[q(l)]$ can be computed using Eq.

(24). The process is then repeated for the next position $q(l+1)$. Note that in each step, $s_{k,i}$ and $u_k$ each have two solutions, but only one solution can be used in the next calculation. For instance, if solution 2 is picked, it must be used consistently throughout all solving processes. When solution 2 of $s_{k,i}$ and $u_k$ are determined for all $q$, $s_{k,i}$ and $u_k$ corresponding to solution 2 are obtained for all wavenumber $k$. $E_u$, $E_{s,i}$, $\Pi_u$ and $\Pi_{s,i}$ related to solution 2 can be solved according to Eqs. (12a-d).

In the numerical computations for a single scalar, e.g., in sections 3.1 and 3.2, both lower-order and higher-order difference methods were applied for better accuracy. In section 3.3, where binary scalar components and two coupling mechanisms are concerned, only the lower-order difference method was applied to reduce computation time. All transitions of the subranges, as frequently observed in the following, are natural variations of the solutions of Eqs. (20), (23), and (24), not artificially determined.

## 3. Results

In this section, we have addressed and resolved several special cases related to the multicomponent model. These cases encompass a range of scenarios. (1) Single scalar component with single coupling mechanism: This subsection examines turbulent systems where there is only one scalar component. For the same problem, Eqs. (18), (20) and (21) are different from that in Zhao's model (Zhao 2022). Therefore, a comparison between the two models is carried out first to ensure the computation results from the two models are consistent. (2) Single scalar component with multiple coupling mechanisms: In this subsection, turbulent systems with a single scalar component but subjected to multiple coupling mechanisms, encompassing diverse forms of interactions or forces, have been analyzed. (3) Binary scalar components with two coupling mechanisms: In this subsection, turbulent systems consist of two scalar components and each corresponds to a different coupling mechanism has been investigated. Our analysis centers around understanding the interactions and coupling between these components, and how they impact the overall cascades of the system. (4) Triple or more scalar components: In this subsection, turbulence systems with three or more scalar components and coupling mechanisms have been preliminarily investigated. Since the features of the inertial subrange have been well established by Kolmogorov, Obukhov, Corrsin and many researchers, in the following section, this investigation only focuses on the features of MFD subranges.

*3.1 A single scalar component with a single coupling mechanism*

Let's begin by considering the simplest case, which involves only a single scalar component and a single coupling mechanism in the model. This scenario is applicable to various models in buoyancy-driven turbulence and EBF-driven turbulence. In buoyancy-driven turbulence, such as thermal convection (Verma et al. 2017), the control scalar is temperature under the Boussinesq approximation. The coupling between the scalar component and the turbulence occurs through a gravity field. In EBF-driven turbulence, for example, EK turbulence, the control scalar can be either electric conductivity or electric permittivity (Zhao & Wang 2017, Zhao & Wang 2019), with the coupling mechanism being an electric field. This simplified scenario captures the essence of these types of turbulent systems, where a single scalar component is influenced by a specific coupling mechanism. Here, $imax = 1$. Considering $M_1 \neq 0$, Eq. (19) becomes

$$\left(\frac{3}{2}k + \frac{5}{2N_1}k^2 s_{k,1}\right)\frac{ds_{k,1}}{dk} + \frac{3\beta_1 - 1}{2}s_{k,1} + \frac{\beta_1 + 1}{2N_1}k s_{k,1}^2 = 0 \qquad (25)$$

This is an alternation of the Eq. (17) in Zhao's model (Zhao 2022). Eq. (25) can be rewritten as

$$\frac{ds_{k,1}}{dk} = -\frac{s_{k,1}}{k}\frac{\frac{3\beta_1-1}{2}+\frac{\beta_1+1}{2N_1}ks_{k,1}}{\frac{3}{2}+\frac{5}{2N_1}ks_{k,1}} \tag{26}$$

From Eq. (26), three different scaling subranges in the MFD subrange can be predicted.

(1) **Constant-$\Pi_u$ subrange**: When $ks_{k,1} \to 0$, Eq. (26) is simplified as $\frac{ds_{k,1}}{dk} = -\frac{s_{k,1}}{k}\frac{3\beta_1-1}{3}$. Thus, $s_{k,1} \sim k^{\frac{1-3\beta_1}{3}}$, and $u_k \sim M_1 k^{-\frac{1}{3}}$ according to Eq. (18). Thereafter, from Eqs. (12), it is simply have $\xi_u = -5/3$, $\xi_{s,1} = -(6\beta_1+1)/3$, $\lambda_u = 0$ and $\lambda_{s,1} = (4-6\beta_1)/3$. This corresponds to the constant-$\Pi_u$ subrange in Zhao's model (Zhao 2022).

(2) **Constant-$\Pi_s$ subrange**: When $ks_{k,1} \to \infty$, Eq. (26) is simplified as $\frac{ds_{k,1}}{dk} = -\frac{s_{k,1}}{k}\frac{\beta_1+1}{5}$. Thus, $s_{k,1} \sim k^{-\frac{\beta_1+1}{5}}$. According to Eq. (18) and Eqs. (12), it is sequentially obtained $u_k \sim M_1 k^{-\frac{2\beta_1-3}{5}}$, $\xi_u = (4\beta_1-11)/5$, $\xi_{s,1} = -(2\beta_1+7)/5$, $\lambda_u = (6\beta_1-4)/5$ and $\lambda_{s,1} = 0$. This corresponds to the constant-$\Pi_s$ subrange in Zhao's model (Zhao 2022).

It should be noted, the region where $ks_{k,1} \to 0$ depends on $\beta_1$. For $\beta_1 < 4/3$, $ks_{k,1} \to 0$ as $k \to 0$, indicating the constant-$\Pi_u$ subrange is at the low-wavenumber limit. For $\beta_1 > 4/3$, $ks_{k,1} \to 0$ as $k \to \infty$, indicating the constant-$\Pi_u$ subrange is at the high-wavenumber limit. For $ks_{k,1} \to \infty$, it is required either $k \to \infty$ for $\beta_1 < 4$, or $k \to 0$ for $\beta_1 > 4$. However, according to Zhao's model (Zhao 2022), $\beta_1 > 4$ leads to a failure of the statistical equilibrium hypothesis, so this situation is not considered in this investigation. Considering the cases together, two different relations between constant-$\Pi_u$ subrange and constant-$\Pi_s$ subrange can be predicted.

For $\beta_1 < 4/3$, the constant-$\Pi_u$ subrange is located on the lower wavenumber side of the constant-$\Pi_s$ subrange. For $4/3 < \beta_1 < 4$, both subranges should be present at the higher wavenumber side, leading to competition between the two subranges, which may depend on $M_1$, $N_1$ and the initial values of $s_{k,1}$. This competition experiences roughly three stages, as numerically investigated by Zhao's model (Zhao 2022). For $4/3 < \beta_1 < 3/2$, both the constant-$\Pi_u$ subrange and constant-$\Pi_s$ subrange are numerically predictable. For $3/2 < \beta_1 < 2$, only constant-$\Pi_u$ subrange is observed. For $2 < \beta_1 < 4$, only the constant-$\Pi_s$ subrange is observed.

(3) **VF subrange**: Beyond inertial subrange, constant-$\Pi_u$ subrange and constant-$\Pi_s$ subrange, there still exists the fourth subrange where $\Pi_u$ and $\Pi_s$ are simultaneously variable, i.e. VF subrange. This is also a special solution for Eq. (26). When $ks_{k,1} = hN_1$ ($h$ is a real number to be determined) which is a limited number irrelevant to $k$, it is required $s_{k,1} = hN_1 k^{-1}$. Eq. (26) becomes

$$\frac{ds_{k,1}}{dk} = -\frac{s_{k,1}}{k}\frac{3\beta_1-1+(\beta_1+1)h}{3+5h} \tag{27}$$

If and only if $h = (3\beta_1-4)/(4-\beta_1)$, Eq. (27) has a solution in the form of $s_{k,1} = hN_1 k^{-1}$. Thus,

$$s_{k,1} = \frac{3\beta_1-4}{4-\beta_1}N_1 k^{-1} \tag{28}$$

According to Eq. (18) and Eqs. (12), it is easy to get the scaling exponents in the VF subrange as

$$\xi_u = \beta_1 - 3, \quad \xi_{s,1} = -3, \quad \lambda_u = \frac{3}{2}\beta_1 - 2, \quad \lambda_{s,1} = \frac{\beta_1}{2} - 2 \tag{29}$$

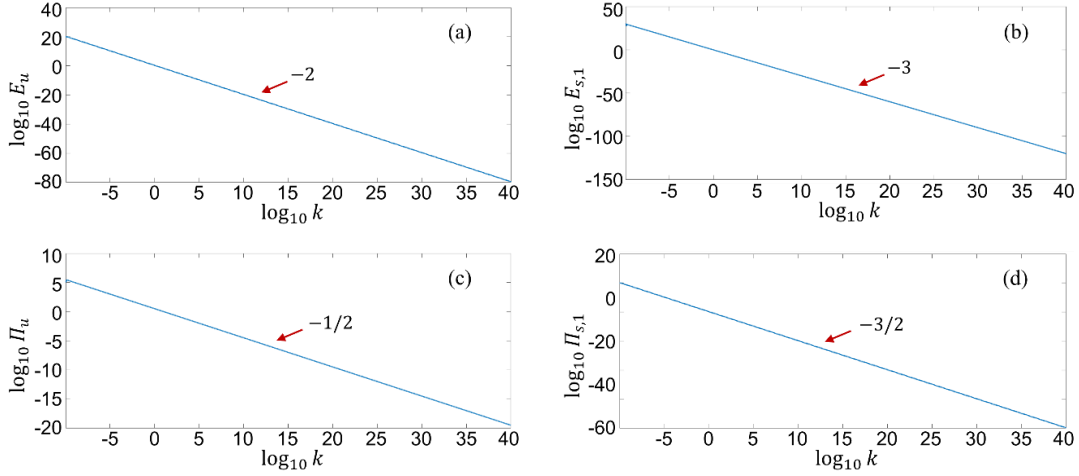

Figure 2. Spectra of kinetic energy, scalar variance and their fluxes for the variable flux subrange when $\beta_1 = 1$. (a) $\log_{10} E_u$ vs $\log_{10} k$, (b) $\log_{10} E_{s,1}$ vs $\log_{10} k$, (c) $\log_{10} \Pi_u$ vs $\log_{10} k$, (d) $\log_{10} \Pi_{s,1}$ vs $\log_{10} k$. It can be seen $\xi_u = -2$, $\xi_{s,1} = -3$, $\lambda_u = -1/2$ and $\lambda_{s,1} = -3/2$ which are also consistent with the prediction in the VF subrange. Please note, the range of $k$ where the scaling properties emerge has no practical meaning. It strictly relies on the selection of $M_1$, $N_1$ and the initial values of $s_{k,1}$. In this case, $M_1 = 1$, $N_1 = 1$, $s_{k,1} = 1$ are used as an example.

Figure 2 presents an example where $\beta_1 = 1$. By numerically solving Eq. (25) using the method detailed in section 2.4, the results $\xi_u = -2$, $\xi_{s,1} = -3$, $\lambda_u = -1/2$ and $\lambda_{s,1} = -3/2$ are obtained, which are coincident with the theoretical predictions. Zhao (2022) provides another example with $\beta_1 = 0$, demonstrating that either $\lambda_u$ or $\lambda_{s,1}$ is non-zero, thus supporting the existence of the VF subrange. The theory of quad-cascade process advanced by Zhao (2022) is further validated through a different theoretical approach. For convenience, the scaling exponents for $\beta_1$ values of 0 and 1 are summarized in Table 1.

When solving Eq. (25), it is important to note the presence of a bifurcation during the transition from the constant-$\Pi_s$ subrange to the VF subrange as $\beta_1$ decreases below 2/3. Figure 3 shows that for $\beta_1 > 2/3$, such as 0.67 and 1, the curves of $E_u$, $E_{s,1}$, $\Pi_u$ and $\Pi_{s,1}$ from the numerical computations are smooth during the transition. However, for

Table 1. Summary of the scaling exponents of different scaling subranges in a single scalar case, where $\beta_1 = 0$ or 1. The scaling exponents are calculated according to Zhao (2022) and Eq. (29) in this investigation.

| $\beta_1$ | Scaling exponents | Inertial subrange | constant-$\Pi_u$ subrange | constant-$\Pi_s$ subrange | VF subrange |
|---|---|---|---|---|---|
| 0 | $\xi_u$ | -5/3 | -5/3 | -11/5 | -3 |
|  | $\xi_{s,i}$ | -5/3 | -1/3 | -7/5 | -3 |
|  | $\lambda_u$ | 0 | 0 | -4/5 | -2 |
|  | $\lambda_{s,i}$ | 0 | -4/3 | 0 | -2 |
| 1 | $\xi_u$ | -5/3 | -5/3 | -7/5 | -2 |
|  | $\xi_{s,i}$ | -5/3 | -7/3 | -9/5 | -3 |
|  | $\lambda_u$ | 0 | 0 | 2/5 | -1/2 |
|  | $\lambda_{s,i}$ | 0 | -2/3 | 0 | -3/2 |

$\beta_1 < 2/3$, even at 0.66 which is slightly smaller than 2/3, spikes with equal intervals (in log coordinates) can be observed in the VF subrange, as shown in Figure 3. These spikes are independent of the computational resolution ($\Delta q$), suggesting the possible existence of other solution branches when solving the inherently nonlinear and inhomogeneous Eq. (25).

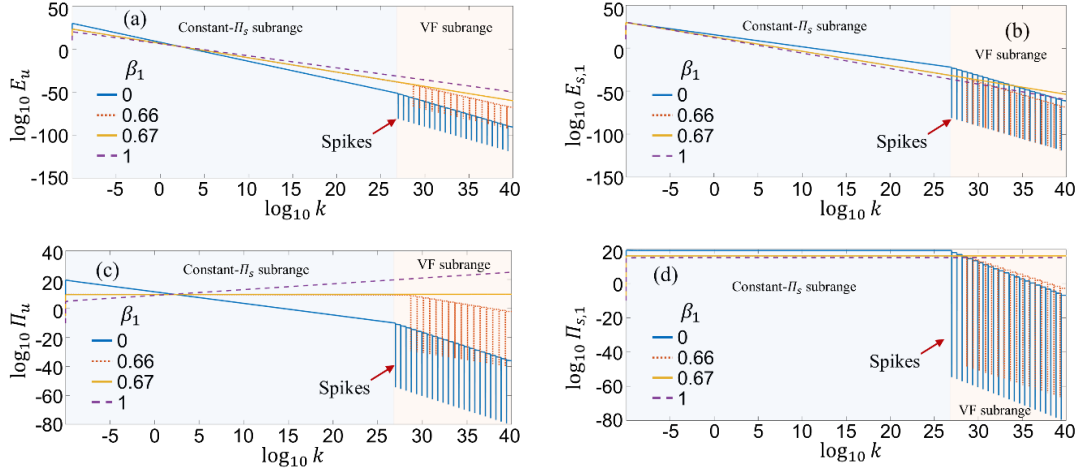

Figure 3. A bifurcation can be observed during the transition from the constant-$\Pi_s$ subrange to the VF subrange, even if in a single scalar component situation when $\beta_1$ is decreased to below 2/3. (a) $\log_{10} E_u$ vs $\log_{10} k$, (b) $\log_{10} E_{s,1}$ vs $\log_{10} k$, (c) $\log_{10} \Pi_u$ vs $\log_{10} k$, (d) $\log_{10} \Pi_{s,1}$ vs $\log_{10} k$.

*3.2 A single scalar component with multiple coupling mechanisms*

Let us consider another scenario where turbulence is driven by multiple coupling mechanisms, all dependent on a single scalar component ($s'$). In this case, the energy feeding rate in Eq. (5c) takes the following general form

$$F_s(\boldsymbol{k}) = \text{Re}[s'(\boldsymbol{k})\boldsymbol{A}(k) \cdot \boldsymbol{u}^*(\boldsymbol{k})] \tag{30}$$

where $\boldsymbol{A}(k) = \sum_{i=1}^{imax} \boldsymbol{M}_i k^{\beta_i}$ is a modulation function which has a polynomial form of $k$, with $\beta_i = i - 1$. Here, the operator of the multiscale force has the form of $\sum_{i=1}^{imax} \boldsymbol{M}_i \mathfrak{D}^{(i-1)/4}$. Since $N_i$ is only a function of $\langle s_i \rangle$, it remains unchanged with $i$ in this case, i.e. $N_i = N$. Besides, let $s_i' = s'$, then Eqs. 3(a) and (b) can be rewritten as

$$\frac{D\boldsymbol{u}}{Dt} = -\frac{1}{\rho}\nabla p + \nu\nabla^2 \boldsymbol{u} + \sum_{i=1}^{imax} \boldsymbol{M}_i \mathfrak{D}^{\frac{(i-1)}{4}} s' \tag{31a}$$

$$\frac{Ds'}{Dt} = -\boldsymbol{N} \cdot \boldsymbol{u} + D_{f,i}\Delta s' \tag{31b}$$

Accordingly, Eqs. (20), (21) and (18) become

$$\left(\frac{3}{2}k + \frac{5}{2N}k^2 s_k\right)\left(\frac{d}{dk}s_k\right)\sum_{i=1}^{imax} M_i^2 + \frac{s_k}{2}\sum_{i=1}^{imax} M_i^2(3i-4) + \frac{ks_k^2}{2N}\sum_{i=1}^{imax} iM_i^2 = 0 \tag{32}$$

$$u_k^2 = \sum_{i=1}^{imax} k^{i-2} M_i s_{k,i} \tag{33}$$

In this section, we discuss two special scenarios to illustrate how multiple coupling mechanisms, even through a single scalar, affect the cascade process of turbulence. One scenario involves turbulence driven by buoyancy and EBF, both caused by variations in the temperature field. The other one considers a modulation function with an exponential form.

**(i) Turbulence driven by buoyancy and EBF**

A typical example is the coexistence of turbulent thermal convection and EK turbulence driven by temperature-dependent electric conductivity. In thermal convection, the flow is driven by temperature fluctuation ($T'$). In EK turbulence, if electric conductivity fluctuation is caused only by temperature fluctuation, then EK turbulence is driven by the temperature field as well. These two mechanisms strongly couple with each other. For single thermal turbulence convection under the Boussinesq approximation, $i = 1$ (or $\beta_1 = 0$) and $\mathbf{M_1} = \alpha g \hat{\mathbf{z}}$, where $\alpha$ and $g$ are the thermal expansion coefficient and gravity acceleration, respectively.

For EK turbulence with small electric conductivity variance, if $\sigma$ is linearly related to temperature fluctuations ($T'$) as $\sigma = \sigma_0 + \frac{\partial \sigma}{\partial T} T'$, with $\sigma_0 = \sigma|_{T=\bar{T}}$, it simply has $\sigma' = \frac{\partial \sigma}{\partial T} T'$. In fluids like water, $\frac{1}{\sigma}\frac{\partial \sigma}{\partial T} = 0.02$ (Ramos et al. 1998, Lide 2009). Considering small scalar fluctuations, i.e. $\sigma' \ll \sigma_0$, it is approximately $\frac{\partial \sigma}{\partial T} \approx 0.02 \sigma_0$, which is constant within the temperature range. For simplicity, define a constant $\Lambda = \frac{\partial \sigma}{\partial T}$. In EK turbulence, it is approximately $i = 2$ ($\beta_2 = 1$). Therefore, $\mathfrak{D}^{\frac{1}{4}} \sigma' = \Lambda \mathfrak{D}^{\frac{1}{4}} T'$. and $\mathbf{M_2} = -\Lambda \frac{\varepsilon E^2 \hat{\mathbf{z}}}{\rho \sigma_0}$.

For this case, $imax = 2$, then Eqs. (31a) and (31b) can be rewritten as

$$\frac{D\mathbf{u}}{Dt} = -\frac{1}{\rho}\nabla p + \nu \nabla^2 \mathbf{u} + \mathbf{M_1} \mathfrak{D}^0 T' + \mathbf{M_2} \mathfrak{D}^{\frac{1}{4}} T' \qquad (34a)$$

$$\frac{DT'}{Dt} = -\mathbf{N_1} \cdot \mathbf{u} + D_{f,1}\Delta T' \qquad (34b)$$

$$\frac{D\sigma'}{Dt} = -\mathbf{N_2} \cdot \mathbf{u} + D_{f,2}\Delta \sigma' \qquad (34c)$$

Here, $\mathbf{N_1} = \nabla \bar{T}$ and $\mathbf{N_2} = \nabla \bar{\sigma} = \Lambda \nabla \bar{T}$. Eq. (34c) becomes

$$\frac{D}{Dt}T' = -\nabla \bar{T} \cdot \mathbf{u} + D_{f,2}\Delta T' \qquad (35)$$

Thus, the problem involves a single scalar $T'$, but two coupling mechanisms. In the MFD subrange, $D_{f,i}$ has negligible influence on the cascade processes of turbulent kinetic energy and scalar variance. According to Eqs. (32) and (33), we obtain

$$\left(\frac{3}{2}k + \frac{5}{2N}k^2 s_k\right)\left(\frac{d}{dk}s_k\right)\left[\alpha^2 g^2 + \Lambda^2 \frac{\varepsilon^2 E^4}{\rho^2 \sigma_0^2}\right] + s_k\left[\Lambda^2 \frac{\varepsilon^2 E^4}{\rho^2 \sigma_0^2} - \frac{\alpha^2 g^2}{2}\right] + \frac{1}{2N}\left[\alpha^2 g^2 + 2\Lambda^2 \frac{\varepsilon^2 E^4}{\rho^2 \sigma_0^2}\right]k s_k^2 = 0 \quad (36)$$

$$u_k^2 = \sum_{i=1}^{2} k^{i-2} M_i s_k = \left(\alpha g k^{-1} - \Lambda \frac{\varepsilon E^2}{\rho \sigma_0}\right) s_k \qquad (37)$$

Assuming the working fluid is water, it has $\sigma_0 = 10^{-3}$ S/m and $\Lambda = 2 \times 10^{-5}$ S/m·K (Ramos et al. 1998, Lide 2009). Let $\alpha = 0.002$ K$^{-1}$ (Putintsev & Putintsev 2017), $g = 9.8$ m/s$^2$, $\rho = 10^3$ kg/m$^3$, $\varepsilon = 7 \times 10^{-10}$ F/m, $E = 10^2$

V/m, $N = 40$ K/m, $M_1 = 0.02$ m/Ks$^2$ and $M_2 = -\Lambda \frac{\varepsilon E^2}{\rho \sigma_0} = \frac{2 \times 10^{-5} \times 7 \times 10^{-10} \times 10^4}{10^3 \times 10^{-3}} = -1.4 \times 10^{-10}$ m$^2$/Ks$^2$. In this case, $M_1 \gg M_2$, indicating the strength of buoyancy-driven turbulence is much higher than that of EK turbulence. The $s_k$ and $u_k$ can be numerically computed through Eqs. (36) and (37). The results are shown in Figure 4 below.

Despite the spikes in solution 1, the entire wavenumber regime can be divided into three subranges from low wavenumber to high wavenumber, including a constant-$\Pi_s$ subrange of buoyancy-driven turbulence, a new VF subrange that has never been reported, and a VF subrange of EK turbulence. In the constant-$\Pi_s$ subrange of buoyancy-driven turbulence, the scaling properties align with BO59 law. The VF subrange corresponding to EK turbulence is also predicted theoretically in Table 1. However, in the new VF subrange, new scaling exponents, i.e. $\xi_u = -6/5$, $\xi_{s,1} = -7/5$, $\lambda_u = 7/10$, $\lambda_{s,1} = 1/2$, have been observed from the numerical computations, as shown in Figures 4(a-d). In solution 2, there are also three subranges emerge in sequence, including constant-$\Pi_u$ and constant-$\Pi_s$ subranges of buoyancy-driven turbulence, and the new VF subrange observed in solution 1.

The emergence of new VF subranges is a consequence of the entanglement of the two mechanisms, even though $M_1 \gg M_2$ indicates the strength of EK turbulence is much smaller than that of buoyancy-driven turbulence. This supports that a secondary mechanism can collaborate with the dominant mechanism to significantly influence the cascade processes. When the wavenumber is sufficiently large, even the much weaker influence of EBF can become dominant.

It is important to note that since $\xi_{s,1} = -7/5$ in the new VF subrange is consistent with the BO59 law, careful analysis is required if a $-7/5$ spectrum is observed in such a flow system. The observed $-7/5$ spectrum may not adhere to the BO59 law but to the VF subrange where buoyancy-driven turbulence and EK turbulence coexist.

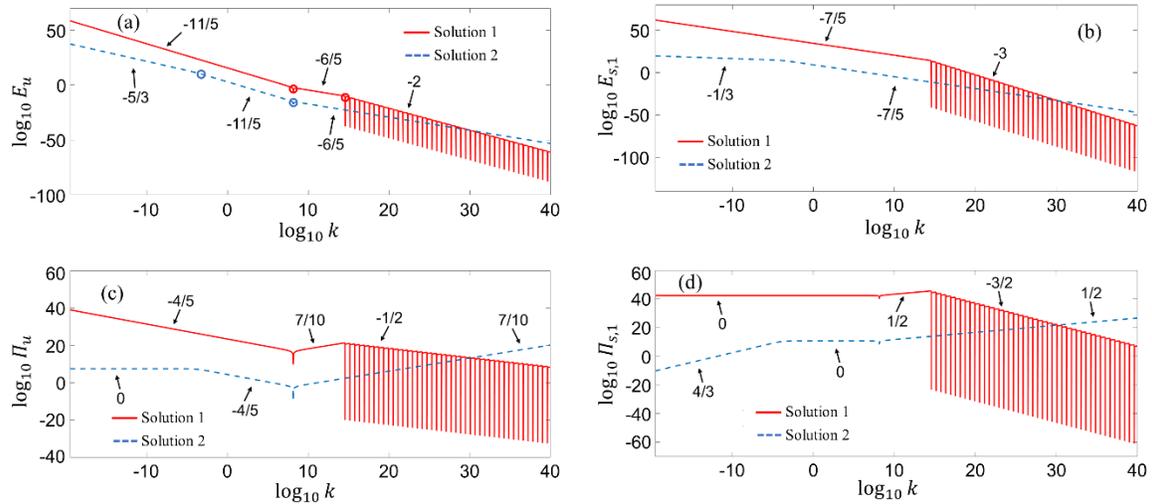

Figure 4. $E_u$, $E_{s,1}$, $\Pi_u$ and $\Pi_{s,1}$ in the turbulence driven by buoyancy and EBF simultaneously. Here, both solution 1 and 2 are taken into account. (a) $\log_{10} E_u$ vs $\log_{10} k$, the red and blue circles represent the intersection positions of different scaling subranges in solution 1 and 2 respectively. (b) $\log_{10} E_{s,1}$ vs $\log_{10} k$. (c) $\log_{10} \Pi_u$ vs $\log_{10} k$. (d) $\log_{10} \Pi_{s,1}$ vs $\log_{10} k$.

In Figure 4(a), the intersection points of the different subranges are plotted. An interesting observation is that the intersection points in solutions 1 and 2 can differ. For instance, in solution 1, the intersection points are located at $k = 10^{8.1}$ and $k = 10^{14.6}$ respectively. In solution 2, the intersection points are located at $k = 10^{-3.2}$ and $k = 10^{8.1}$ respectively. The various intersection points of the different subranges indicate more characteristic scales in turbulence driven by buoyancy and EBF, even though the latter is much weaker than the former.

**(ii) Modulated by an exponential function**

In recent years, several investigations have reported on the modulations directly applied to turbulence (DeDominicis & Martin 1979, Bhattacharjee 2022). However, in momentum-scalar coupling turbulence, the coexistence of multiple coupling mechanisms introduces a more complex modulation on the spectra of turbulent kinetic energy and scalar variance. Using the modulation form in Eq. (30), a general modulation model has been established. For instance, if the modulation function takes an exponential form, $A(k) = A_0 e^{-k/C}$ (with $C$ being a reference wavenumber), it can be approximated as $A(k) \approx A_0 \left[1 - \frac{k}{C} + \frac{1}{2}\left(\frac{k}{C}\right)^2 - \frac{1}{6}\left(\frac{k}{C}\right)^3\right] + O(k^4)$ using Taylor expansion when $k/C \ll 1$. This approximation is equivalent to having $\beta_i = 0, 1, 2, 3$, with $M_i$ being $1, -\frac{1}{C}, \frac{1}{2C^2}$ and $-\frac{1}{6C^3}$ respectively. Subsequently, $s_k$ and $u_k$ can be solved from Eqs. (32) and (33), as shown in Figure 5. Here, only one solution is presented.

In the first case, with $A_0 = 1$ and $C = 1$, the $M_i$ values are 1, -1, 1/2, and -1/6 respectively. The small differences among $M_i$ indicate similar forcing intensities among these coupling mechanisms. This scenario demonstrates how multiple mechanisms, rather than a single one, control the cascade process in momentum-scalar coupling turbulence. The cascade process is divided into two new VF subranges intersecting at a wavenumber $k_i = 1.596$, where a singular point is observed in Figure 5(a), (c), and (d). In the VF subrange at the lower wavenumber regime, the scaling exponents are $\xi_u = -59/25$, $\xi_{s,1} = -17/10$, $\lambda_u = -1$, and $\lambda_{s,1} = -1/3$. In the VF subrange at the higher wavenumber regime, the scaling exponents are $\xi_u = 2/3$, $\xi_{s,1} = -17/10$, $\lambda_u = 88/25$ and $\lambda_{s,1} = 34/33$. These scaling exponents, not predicted in any single scalar component model, strongly depend on the form of $\boldsymbol{A}(k)$, indicating that nonlinear interactions in regimes with multiple coupling mechanisms can lead to unexpected scaling properties. It should be noted that for $k/C > 1$, it is inappropriate to describe the influence of $\boldsymbol{A}(k) = A_0 e^{-k/C}$ modulation using the approximation $\boldsymbol{A}(k) \approx A_0 \left[1 - \frac{k}{C} + \frac{1}{2}\left(\frac{k}{C}\right)^2 - \frac{1}{6}\left(\frac{k}{C}\right)^3\right]$.

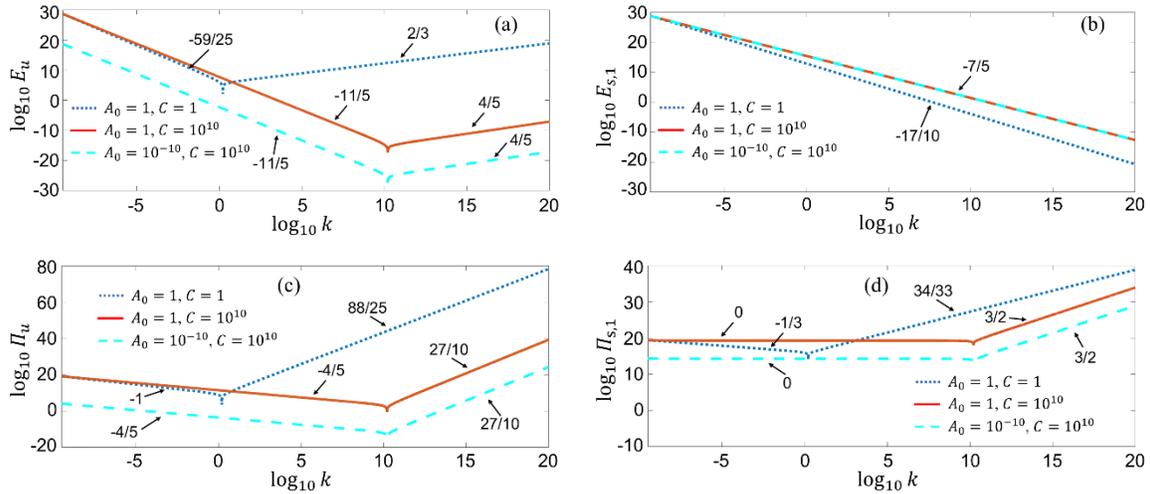

Figure 5. $E_u$, $E_{s,1}$, $\Pi_u$ and $\Pi_{s,1}$ in the turbulence driven by a single scalar with multiple coupling mechanisms. The modulation function has a form of $A(k) = A_0 e^{-k/C} \approx A_0 \left[1 - \frac{k}{C} + \frac{1}{2}\left(\frac{k}{C}\right)^2 - \frac{1}{6}\left(\frac{k}{C}\right)^3\right]$. Here, only solution 1 is computed. (a) $\log_{10} E_u$ vs $\log_{10} k$, (b) $\log_{10} E_{s,1}$ vs $\log_{10} k$, (c) $\log_{10} \Pi_u$ vs $\log_{10} k$, (d) $\log_{10} \Pi_{s,1}$ vs $\log_{10} k$.

In the second case, with $A_0 = 1$ but $C$ increased to $10^{10}$, the larger $C$ significantly inhibits the contributions of forcing mechanisms with larger $\beta_i$. Consequently, the influence of $\beta_1 = 0$ is promoted, controlling the cascade process. The constant-$\Pi_s$ subrange and VF subrange of $\beta_1 = 0$ are observed sequentially. Additionally, $k_i$ is postponed to $1.596 \times 10^{10}$ due to the larger $C$, indicating a tight relationship between $k_i$ and $M_i$.

In the third case, where $A_0$ is changed to $10^{-10}$ but $C$ is kept at $10^{10}$, the cascade process remains the same as in the previous case, differing only in the smaller magnitudes of $E_u$, $\Pi_u$ and $\Pi_{s,1}$ due to the smaller $u_k$ (see Eq. (37)). The magnitude of $A_0$ is irrelevant to $E_{s,1}$.

### *3.3 Binary scalar components and two coupling mechanisms*

In the study of momentum-scalar coupling turbulence, we often encounter scenarios where multiple scalar components and mechanisms can coexist simultaneously. Among these scenarios, binary scalar transport stands out as both a simple and representative example of the complexities involved in multiple scalar transport. This section delves into the cascades of turbulent kinetic energy and scalar variance, driven by binary scalar components that operate independently of each other. This situation is commonly observed in both natural environments and engineering applications.

A notable illustration of this is the simultaneous occurrence of stratified turbulence and EK turbulence. These phenomena are driven by buoyancy and EBF, respectively. The buoyancy force depends on density variations, whereas the EBF is influenced by electric conductivity or permittivity. In instances where density fluctuations are not substantial enough to provoke notable variations in electric conductivity or permittivity, and vice versa, it is logical to treat these quantities as independent from one another. Consequently, density and electric conductivity (or electric permittivity) function as independent scalar quantities, denoted as $s_{k,1}$ and $s_{k,2}$ respectively. These scalar quantities can be addressed and solved using Eq. (18) along with Eqs. (20-22), considering a scenario with $imax = 2$, as

$$M_1^2 \left[\left(\frac{3}{2}k + \frac{5}{2N_1}k^2 s_{k,1}\right)\left(\frac{d}{dk} s_{k,1}\right) + \frac{1}{2}(3\beta_1 - 1)s_{k,1} + \frac{\beta_1 + 1}{2N_1} k s_{k,1}^2\right]$$
$$+ M_2^2 \left[\left(\frac{3}{2}k + \frac{5}{2N_2}k^2 s_{k,2}\right)\left(\frac{d}{dk} s_{k,2}\right) + \frac{1}{2}(3\beta_2 - 1)s_{k,2} + \frac{\beta_2 + 1}{2N_2} k s_{k,2}^2\right]$$
$$+ 2M_1 M_2 \left[\left(\frac{3}{2}k + \frac{2}{N_2}k^2 s_{k,2} + \frac{1}{2N_1}k^2 s_{k,1}\right)\left(\frac{d}{dk} s_{k,2}\right) + \frac{1}{2}(3\beta_2 - 1)s_{k,2} + \frac{\beta_2 + 1}{2N_1} k s_{k,1} s_{k,2}\right] = 0$$
(38)

$$\frac{3}{2}k \frac{d}{dk}(s_{k,2} - s_{k,1}) + k^2 \frac{d}{dk}\left(\frac{s_{k,2}^2}{N_2} - \frac{s_{k,1}^2}{N_1}\right) + \left(\frac{1}{2N_1} k^2 s_{k,1} \frac{d}{dk} s_{k,2} - \frac{1}{2N_2} k^2 s_{k,2} \frac{d}{dk} s_{k,1}\right)$$
$$+ \frac{3}{2}(\beta_2 s_{k,2} - \beta_1 s_{k,1}) - \frac{1}{2}(s_{k,2} - s_{k,1}) + \frac{k s_{k,1} s_{k,2}}{2N_1 N_2}(\beta_2 N_2 + N_2 - \beta_1 N_1 - N_1) = 0 \quad (39)$$

$$k u_k^2 = k^{\beta_1} M_1 s_{k,1} + k^{\beta_2} M_2 s_{k,2} \quad (40)$$

In the following sections, some examples of the cascade processes due to binary scalar components will be discussed.

**(i) A special scenario where $\beta_1 = \beta_2$, $M_1 = M_2$, $D_{f,1} = D_{f,2}$, but $N_1 \neq N_2$**

In this specific scenario, we explore a straightforward example of binary scalar transport, where the two scalar components can be different, yet the underlying driving mechanism across these scalars remains identical. For instance, in EK turbulence, two variants of electric conductivity components (such as Ca$^{2+}$ and Cl$^-$) with distinct $N_1$ and $N_2$ values are incorporated. This results in the equations being approximately defined as $\beta_1 = \beta_2$, $M_1 = M_2$, and $D_{f,1} = D_{f,2}$, although $N_1 \neq N_2$. Consequently, under these assumptions, Eqs. (3b) are reformulated as

$$\begin{cases} \dfrac{Ds_1'}{Dt} = -\mathbf{N_1} \cdot \mathbf{u} + D_{f,1}\Delta s_1' \\ \dfrac{Ds_2'}{Dt} = -\mathbf{N_2} \cdot \mathbf{u} + D_{f,1}\Delta s_2' \end{cases} \tag{41}$$

Any linear combination of $s_1'$ and $s_2'$, defined as $s_c' = as_1' + bs_2'$ ($a$ and $b$ are two proportion coefficients), also fulfills a convection-diffusion equation

$$\frac{Ds_c'}{Dt} = -\mathbf{N_c} \cdot \mathbf{u} + D_{f,1}\Delta s_c' \tag{42}$$

where $\mathbf{N_c} = a\mathbf{N_1} + b\mathbf{N_2}$. Therefore, theoretically, for the special binary scalar scenario discussed in this section, analyzing $s_1'$ and $s_2'$ should yield results identical to those found in section 3.1 for a single scalar, $s_c'$. Taking $a = b = 1$ as an example, within the context of a single scalar model and referring to section 3.1, Eq. (25) can be rewritten as

$$\left(\frac{3}{2}k + \frac{5}{2N_c}k^2 s_{k,c}\right)\frac{ds_{k,c}}{dk} + \frac{3\beta_1 - 1}{2}s_{k,c} + \frac{\beta_1 + 1}{2N_c}ks_{k,c}^2 = 0 \tag{43}$$

where $s_{k,c} = s_{k,1} + s_{k,2}$ is the spectral component of $s_c'$. After substituting $s_{k,c}$ and $N_c$ into Eq. (43), it becomes

$$\frac{3}{2}k\frac{d(s_{k,1}+s_{k,2})}{dk} + \frac{5}{2(N_1+N_2)}k^2 s_{k,1}\frac{ds_{k,1}}{dk} + \frac{5}{2(N_1+N_2)}k^2 s_{k,2}\frac{ds_{k,2}}{dk} + \frac{5}{2(N_1+N_2)}k^2 s_{k,1}\frac{ds_{k,2}}{dk}$$
$$+ \frac{5}{2(N_1+N_2)}k^2 s_{k,2}\frac{ds_{k,1}}{dk} + \frac{3\beta_1-1}{2}(s_{k,1}+s_{k,2}) + \frac{\beta_1+1}{2(N_1+N_2)}k(s_{k,1}^2 + s_{k,2}^2 + 2s_{k,1}s_{k,2}) = 0 \tag{44}$$

When considering binary scalar components $s_1'$ and $s_2'$, according to Eq. (38), it is obtained

$$3k\frac{d}{dk}(s_{k,1}+s_{k,2}) + \frac{9}{2N_1}k^2 s_{k,1}\frac{d}{dk}s_{k,1} + \frac{9}{2N_2}k^2 s_{k,2}\frac{d}{dk}s_{k,2} + \frac{1}{2N_1}k^2 s_{k,1}\frac{d}{dk}s_{k,2} + \frac{1}{2N_2}k^2 s_{k,2}\frac{d}{dk}s_{k,1}$$
$$+(3\beta_1-1)(s_{k,1}+s_{k,2}) + (\beta_1+1)k\left(\frac{s_{k,1}^2}{2N_1} + \frac{s_{k,2}^2}{2N_2}\right) + (\beta_1+1)\left(\frac{N_1+N_2}{2N_1N_2}\right)ks_{k,1}s_{k,2} = 0 \tag{45}$$

Both Eqs. (44) and (45) describe the same phenomenon, indicating they must be either equivalent or proportional. To satisfy this, we introduce the assumption that $s_{k,1}$ and $s_{k,2}$ share the same scaling behavior, differing only by a factor, $h_c$, such that $s_{k,2} = h_c s_{k,1}$. Consequently, Eqs. (44) and (45) are equivalent if $h_c = N_2/N_1$. This leads us to the following relationship

$$s_{k,2} = \frac{N_2}{N_1}s_{k,1} \tag{46}$$

In this specific scenario, one can solve for $s_{k,c}$ first. Afterwards, $s_{k,1}$ and $s_{k,2}$ can be calculated using $s_{k,1} = \frac{N_1}{N_1+N_2}s_{k,c}$ and $s_{k,2} = \frac{N_2}{N_1+N_2}s_{k,c}$, respectively. Alternatively, the same results can also be derived by solving Eqs. (39) and (45) directly. Once $s_{k,1}$ and $s_{k,2}$ are obtained, $u_k$ can be calculated through Eq. (40).

Figures 6(a) and 6(b) illustrate further insights. Figure 6(a) showcases the calculated $E_{s,i}$ for $\beta_1 = 1$ and $N_2 = 10N_1$, according to Eqs. (39) and (45). Despite $N_2$ being an order of magnitude larger than $N_1$, $E_{s,1}$ and $E_{s,2}$ exhibit remarkably similar distributions, with $E_{s,2} = 10^2 E_{s,1}$ within the constant-$\Pi_u$ subrange (as shown in the inset of Figure 6(a)). In Figure 6(b), the spectra of $s_1'$, $s_2'$, and $s_c'$ are plotted, derived from Eqs. (45) and (39), and Eq. (43), respectively. Initial values, denoted by circles, reveal $E_{s,c} = E_{s,1} + E_{s,2}$, translating to $s_{k,c} = 11 s_{k,1}$ and $s_{k,c} = \frac{11}{10} s_{k,2}$, aligning with the theoretical expectations.

This analysis demonstrates that the discussed special scenario can be resolved using either a single or binary scalar component model approach. Furthermore, the model can be extended to incorporate multiple scalar components (for instance, when $imax > 2$), assuming $\beta_i$, $M_i$ and $D_{f,i}$ remain constant across all components, while $N_i$ varies. Thereafter,

$$s_{k,i} = \frac{N_i}{\sum_i N_i} s_{k,c} \quad \text{and} \quad s_{k,c} = \sum_i s_{k,i} \tag{47}$$

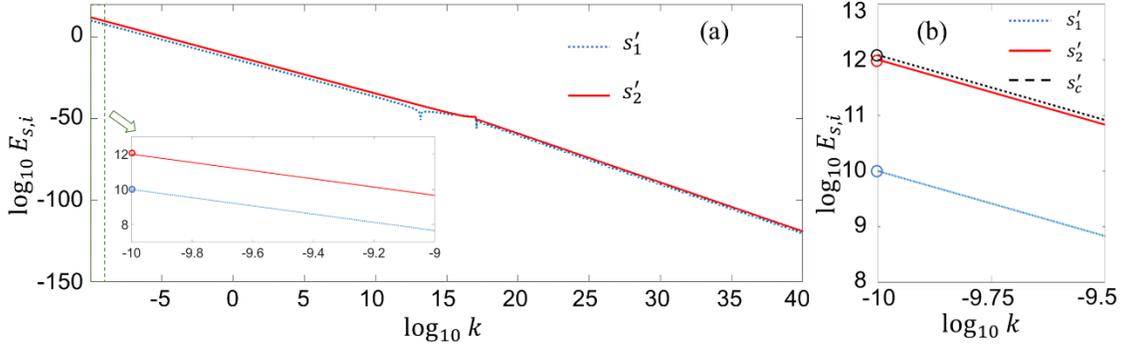

Figure 6. Spectra of scalar variance at $\beta_1 = 1$. Here, $N_2 = 10N_1$. (a) $\log_{10} E_{s,1}$ and $\log_{10} E_{s,2}$ vary with $\log_{10} k$. $E_{s,1}$ and $E_{s,2}$ are computed from Eqs. (45) and (39). The inset is the zoom-in of the plots. The circles represent the initial values of $E_{s,1}$ and $E_{s,2}$ at the lowest wavenumber. (b) Comparison among the spectra of $s_1'$ and $s_2'$ computed from Eqs. (45) and (39), and $s_c'$ computed from Eq. (43). The circles represent the initial values of $E_{s,1}$, $E_{s,2}$ and $E_{s,c}$ at the lowest wavenumber.

**(ii) General scenario of binary scalar components**

In the general scenario involving two binary scalar components, Eqs. (38) and (39) must be solved concurrently to determine $s_{k,1}$ and $s_{k,2}$. Subsequently, $u_k$ is calculated using Eq. (40). Following this, $E_u$, $E_{s,i}$ ($i = 1,2$), $\Pi_u$ and $\Pi_{s,i}$ can be derived from $s_{k,1}$, $s_{k,2}$ and $u_k$ directly, based on Eqs. (12).

Previous studies have shown that in the wavenumber space, the influence of buoyancy or EBF can be differentiated; buoyancy predominantly affects the cascades in the lower wavenumber ranges, while EBF govern the higher wavenumber ranges. However, the interaction between these mechanisms in a regime where both are present remains unclear. This section explores the combined effects of buoyancy ($\beta_1 = 0$) and EBF ($\beta_2 = 1$) with varying strengths on the cascades of turbulent kinetic energy and scalar variance. The two scalar components considered could represent temperature and electric conductivity, respectively.

Figure 7 illustrates the wavenumber subranges influenced by both buoyancy and EBF. In a conceptual model setting, the magnitudes of $M_i$ and $N_i$ were arbitrarily chosen. With $M_1 = M_2 = 1$ and $N_1 = N_2 = 1$ as an example, solution 1 (depicted by red lines) shows a sequence of a constant-$\Pi_u$ subrange followed by two VF subranges. In the constant-$\Pi_u$ subrange, located in the lower wavenumber regime, $\xi_u = -5/3$ (Figure 7(a)) and $\lambda_u = 0$ (Figure 7(b)), $\xi_{s,1} = -1/3$ (Figure 7(c)), $\xi_{s,2} = -7/3$ (Figure 7(d)), $\lambda_{s,1} = 4/3$ (Figure 7(e)) and $\lambda_{s,2} = -2/3$ (Figure 7(f)) which are aligning with single scalar predictions in Table 1 for $\beta_1 = 0$ and $\beta_2 = 1$, respectively.

Following the constant-$\Pi_u$ subrange is a VF subrange driven by buoyancy, where $\xi_u = -3$, $\xi_{s,1} = \xi_{s,2} = -3$, $\lambda_u = -2$, and $\lambda_{s,1} = \lambda_{s,2} = -2$. In this subrange, EBF plays a negligible role, and the cascades of $s'_2$ are the same as those of $s'_1$. Proceeding to the higher wavenumber side, another VF subrange completely governed by EBF is visible, where the influence of buoyancy is negligible, and $\xi_u = -2$, $\xi_{s,1} = \xi_{s,2} = -3$, $\lambda_u = -1/2$, $\lambda_{s,1} = \lambda_{s,2} = -3/2$ which are coincident to Table 1 as well.

When considering a scenario where the second mechanism is weaker ($M_2 = 10^{-3}$ and $N_2 = 10^{-3}$), solution 1 (depicted by black dashed lines) presents two distinct subranges. One is the constant-$\Pi_u$ subrange located at the lower wavenumber regime attributed to buoyancy, and the other is the VF subrange located at the higher wavenumber regime attributed to EBF. According to the unbalanced forcing through buoyancy and EBF, the VF subrange of buoyancy-driven turbulence mentioned in last paragraph is not observed in this case. Instead, a highly fluctuated transitional regime is present, where singular and intermittent (sudden fall) spectra are numerically computed (see e.g. Figure 7(b), (d), (e) and (f)). Despite the reduced magnitudes of $M_2$ and $N_2$, their effect on the overall scaling in the constant-$\Pi_u$ subrange dominated by buoyancy, remains unchanged. In the VF subrange dominated by EBF, the values of $E_u$, $E_{s,i}$ ($i=1,2$), $\Pi_u$ and $\Pi_{s,i}$ are all significantly lower than those in the previous scenario, as illustrated in Figure 7. Yet, an intriguing exception is observed: smaller values of $M_2$ and $N_2$ lead to a reduction in $\Pi_{s,1}$ within this subrange, without impacting $E_{s,1}$. Such a phenomenon is not observed in conventional hydrodynamic turbulence. Nonetheless, in turbulence that involves momentum-scalar coupling, this behavior may be more common. This is because $\Pi_{s,i}$ (as defined in Eq. (12d)) are more influenced by the smaller $M_2$, mediated through $u_k$ (Eq. (18)).

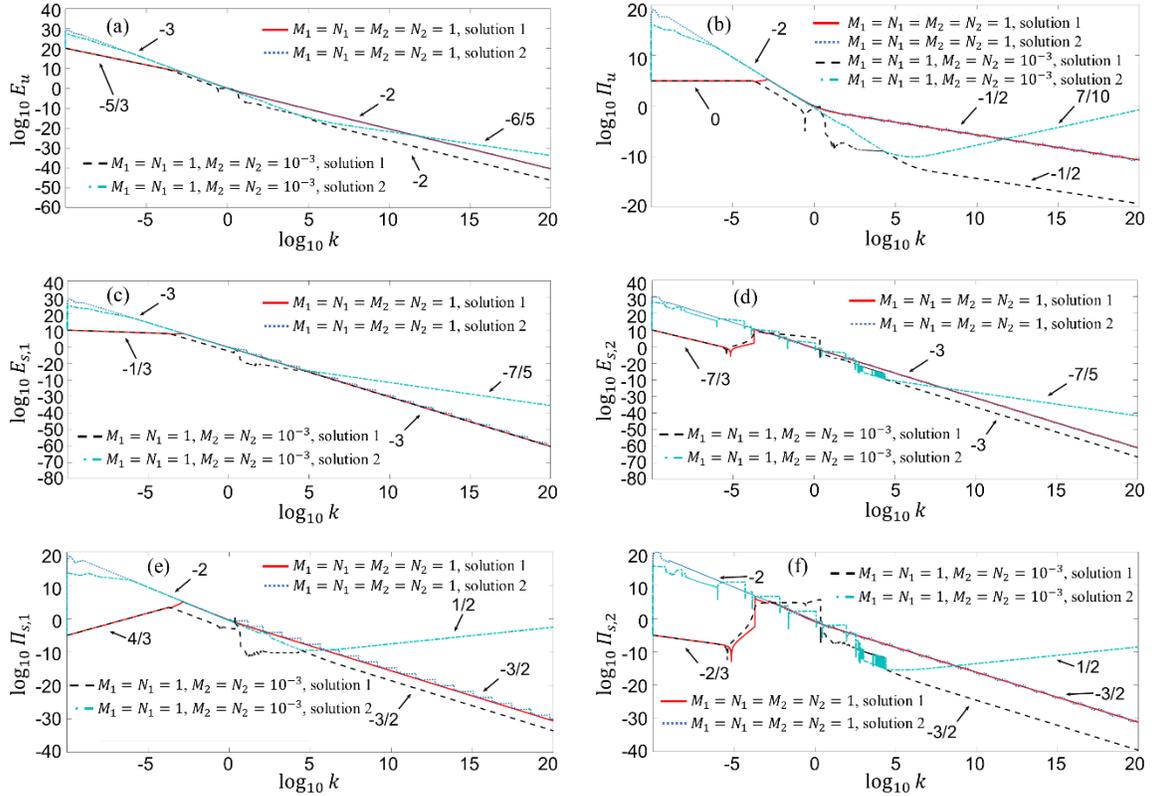

Figure 7. Spectra computed with binary scalar components at different $\beta_i$. Here, $\beta_1 = 0$ and $\beta_2 = 1$ to simulate the turbulence driven by buoyancy and EBF simultaneously. As a conceptual model, the magnitude of $M_i$ and $N_i$ are selected arbitrarily. (a) $\log_{10} E_u$ vs $\log_{10} k$, (b) $\log_{10} \Pi_u$ vs $\log_{10} k$, (c) $\log_{10} E_{s,1}$ vs $\log_{10} k$, (d) $\log_{10} E_{s,2}$ vs $\log_{10} k$, (e) $\log_{10} \Pi_{s,1}$ vs $\log_{10} k$, (f) $\log_{10} \Pi_{s,2}$ vs $\log_{10} k$.

The impact of solution 2 on the previously mentioned scenarios has also been examined, with findings presented in Figure 7. Analyzing the spectra indicated by blue dotted lines, when $M_1 = M_2 = 1$ and $N_1 = N_2 = 1$, the cascade divides into two distinct subranges. The first is a VF subrange of buoyancy-driven turbulence found in the lower wavenumber regime. The second subrange, situated in the higher wavenumber region, pertains to EK turbulence. Despite the appearance of a "zigzag" pattern in this higher subrange (as seen in parts (b), (e), and (f) of Figure 7), the overall trend aligns with the predictions for the VF subrange of EK turbulence detailed in Table 1.

Conversely, when implementing solution 2 with $M_1 = 1$, $N_1 = 1$, $M_2 = 10^{-3}$ and $N_2 = 10^{-3}$ (represented by blue dot-dashed lines), novel scaling characteristics emerge. In the lower wavenumber regime, the VF subrange for buoyancy-driven turbulence remains noticeable. Yet, within the higher wavenumber sphere, the new VF subrange—previously discussed in section 3.2—emerges again, characterized by $\xi_u = -6/5$, $\xi_{s,1} = \xi_{s,2} = -7/5$, $\lambda_u = 7/10$, and $\lambda_{s,1} = \lambda_{s,2} = 1/2$.

These observations underscore the nuanced interplay between buoyancy and EBF in turbulence, suggesting that a variety of scaling behaviors can emerge based on the relative strengths of these mechanisms. This complexity invites further investigation, particularly in scenarios with more varied magnitudes of $M_i$ and $N_i$, indicating the potential for uncovering additional unpredictable scalings in momentum-scalar coupling turbulence.

*3.4 Triple or more scalar components*

When there are three or more independent scalar components, that is, $imax \geq 3$, solving for the $imax + 1$ unknowns ($u_k$ and $s_{k,i}$) using the given equations ($(imax^2 - imax)/2 + 2$) from Eqs. (18), (20), and (21) leads to a nonclosure issue. The problem appears overdetermined at first glance because there are more equations than unknowns, making the determination of the solutions exceedingly challenging. However, in a physical system, such as EK turbulence involving more than three electrolyte components (e.g., $Na^+$, $K^+$, $Ca^{2+}$, $Mg^{2+}$, $Cl^-$, and others), deterministic solutions for $u_k$ and $s_{k,i}$ should be existed and experimentally measurable. This suggests that, in reality, the solutions should be unique and attainable. Therefore, it can be deduced that there are $(imax^2 - 3imax)/2 + 1$ constraints necessary to make the equations solvable.

Identifying these unknown constraints is vital for comprehending complex nonlinear systems with multiple components and coupling mechanisms, which are more applicable to real-world situations than the highly simplified models like K41 (Kolmogorov 1941), Bolgiano-Obukhov (Bolgiano 1959, Obukhov 1959), and Zhao-Wang model (Zhao & Wang 2017) etc. A prime example is in electrochemical engineering, where the mixture of complex chemical components in a turbulent mixer with the presence of EK flow facilitates chemical reactions. Knowing the exact spatial or spectral distribution of each chemical component is crucial to predicting the reaction outcomes. Nevertheless, this topic exceeds the current study's scope and will be addressed in future research.

# 4. Conclusions

This study introduces a theoretical model for the cascade processes in momentum-scaling coupling turbulence driven by multiple scalar components and various mechanisms, building upon the general flux model proposed by Zhao (2022). The current model comprehensively supports the quad-cascade processes of turbulent kinetic energy and scalar variance, while also forecasting new scaling properties.

For a model with a single scalar component where only one coupling mechanism is taken into account, it predicts that the variable fluxes subrange can exhibit scaling exponents $\xi_u = \beta_1 - 3$, $\xi_{s,1} = -3$, $\lambda_u = \frac{3}{2}\beta_1 - 2$ and $\lambda_{s,1} = \frac{1}{2}\beta_1 - 2$, derived from the order of deviation in the multiscale force $\boldsymbol{M_i}\mathfrak{D}^{\beta_i/4}s_i'$. Remarkably, spikes are identified in the variable fluxes subranges when $\beta_i < 2/3$, suggesting the emergence of local intermittency when solving the inherently nonlinear and inhomogeneous conservation equation.

In the context of turbulence driven by a single scalar but subjected to multiple coupling mechanisms, for instance, the combined effects of buoyancy-driven turbulence and EK turbulence due to temperature variations, a distinctive variable fluxes subrange is detected. This subrange exhibits scaling exponents $\xi_u = -6/5$, $\xi_{s,1} = -7/5$, $\lambda_u = 7/10$ and $\lambda_{s,1} = 1/2$. With the expansion of this model to include an exponential modulation function, the emergence of two new variable fluxes subranges is disclosed. The first, in the lower wavenumber regime, displays scaling exponents $\xi_u = -59/25$, $\xi_{s,1} = -17/10$, $\lambda_u = -1$ and $\lambda_{s,1} = -1/3$. The second appears in the higher wavenumber regime, marked by scaling exponents $\xi_u = 2/3$, $\xi_{s,1} = -17/10$, $\lambda_u = 88/25$ and $\lambda_{s,1} = 34/33$. These findings reveal that the determined scaling exponents heavily rely on the chosen modulation function. Moreover, it hints at the potential to manipulate the scaling properties and cascade processes by strategically aligning different scalar components and coupling mechanisms.

Subsequently, this study delves into a binary scalar components model, specifically focusing on electrokinetic turbulence ($\beta_i = 1$) driven by two types of ions with disparate mean concentration gradients. Two approaches are compared: one using a single scalar component model based on the linear relationship between the two scalar transport equations, and the other directly computing the cascade process using the binary scalar component model. These distinct approaches both converge on the same outcomes regarding the cascade processes, indicating consistency in the results derived from simplified and more complex models.

Further exploration into the interplay between buoyancy-driven turbulence ($\beta_1 = 0$) and electrokinetic turbulence ($\beta_2 = 1$), characterized by temperature and electric conductivity as the two scalar components, reveals that the cascade process is comprised of several distinct subranges, including the constant-$\Pi_u$ subrange, VF subranges of buoyancy-driven turbulence and EK turbulence etc. Notably, the new variable fluxes subrange characterized by $\xi_u = -6/5$, $\xi_{s,1} = \xi_{s,2} = -7/5$, $\lambda_u = 7/10$, $\lambda_{s,1} = \lambda_{s,2} = 1/2$ is observed again. This finding is further corroborated by using the single scalar component model influenced by dual mechanisms.

The investigation then addresses the complexities that arise when three or more distinct scalar components and coupling mechanisms simultaneously coexist, leading to potentially overdetermined conservative equations. Yet, from a physical standpoint, it is posited that unique solutions for the cascades of multiple scalars, which can be empirically validated, must exist. It suggests that additional equations or constraints are needed to circumvent these complexities, setting the stage for further research endeavors.

Turbulence is a complex and challenging phenomenon, particularly in the context of momentum-scalar coupling, where multiple scalar components and coupling mechanisms come into play. Despite its importance, the cascade process of momentum-scalar coupling turbulence has received limited attention and remains poorly understood. In this study, we have undertaken an effort to address this knowledge gap and shed light on a small part of this complex problem, which is bound to be confronted sooner or later. By doing so, we aim to uncover the diverse scaling properties observed experimentally in momentum-scalar coupling turbulence, including phenomena in buoyancy-driven turbulence in atmosphere. Understanding the cascade processes and scaling properties in these turbulent flows is crucial for various scientific fields, not only for the conventional hydrodynamic turbulence, but also for theoretical physics and optics (Tang et al. 2020, Al-Mahmoud et al. 2022), particularly in wave turbulence (Nazarenko 2011, Zhu et al. 2023) when more complex coupling of multiple wave functions must be included. Our findings contribute to the groundwork necessary for future investigations and deepen our understanding of turbulent flows across different physical systems.

**Acknowledgement** This investigation is supported by National Natural Science Foundation of China (Grant No. 51927804)
**Declaration of Interests.** The authors report no conflict of interest.